\documentstyle[preprint,aps]{revtex}

\draft
\newcommand{\be}{\begin{equation}}
\newcommand{\ee}{\end{equation}}
\newcommand{\ben}{\begin{eqnarray}}
\newcommand{\een}{\end{eqnarray}}
\newcommand{\ra}{\rangle}
\newcommand{\la}{\langle}

\newcommand{\iii}{\'{\i}}

\begin{document}
\draft
\title{On the ``Fake" Inferred Entanglement Associated with the
Maximum Entropy Inference of Quantum States}
\author{J. Batle$^1$, M. Casas$^1$, A. R. Plastino$^{1,\,2,\,3}$,
and A. Plastino$^{3,\,4}$}

\address {
$^1$Departament de F\iii sica, Universitat de les Illes Balears,
07071 Palma de Mallorca, Spain \\
$^2$Faculty of Astronomy and Geophysics, National University La Plata,
  C.C. 727, 1900 La Plata \\
  $^3$Argentina's National Research Council (CONICET) \\
$^4$Department of Physics, National University La Plata,
  C.C. 727, 1900 La Plata, Argentina
}

\date{\today}

\maketitle

\begin{abstract}
The inference of entangled quantum states by
recourse to the maximum entropy principle is
considered in connection with the recently pointed
out problem of fake inferred entanglement [R.
Horodecki, {\it et al.}, Phys. Rev. A {\it 59}
(1999) 1799]. We show that there are operators $\hat A$,
both diagonal and non diagonal in the Bell basis, such
that when the expectation value $\langle \hat A \rangle$
is taken as prior information the problem of fake
entanglement is not solved by adding a new constraint
associated with the mean value of $\hat A^2$ (unlike
what happens when the partial information is given by
the expectation value of a Bell operator). The fake
entanglement generated by the maximum entropy
principle is also studied quantitatively by
comparing the entanglement of formation of the
inferred state with that of the original one.

\vskip 5mm
 Pacs: 03.67.-a; 89.70.+c;
03.65.Bz
\end{abstract}
\vspace{.5cm}


\noindent \hskip 2cm
Quantum Information Theory

\section{Introduction}
 The inference of entangled quantum states by recourse to the maximum entropy
 principle has been recently considered in the literature
 \cite{HHH99,BDADK97,RPPC00,R99,AR99}. In particular, the question of how to
estimate in a reliable way  the amount of entanglement of a bipartite
quantum system when only partial, incomplete information about its state
is available was addressed by Horodecki {\it et al.} \cite{HHH99}. Various
strategies have been advanced in order to tackle this problem
\cite{HHH99,RPPC00,R99,AR99,SH00}. Horodecki's question has also been
considered in connection with procedures for the entanglement purification
of unknown quantum states \cite{BCS00}. The motivation behind these lines
of inquiry is that quantum entanglement is the basic resource required to
implement several of the most important processes studied by quantum
information theory \cite{LPS98,WC97,W98}, such as quantum
cryptographic key distribution \cite{E91}, quantum teleportation
\cite{BBCJPW93}, superdense coding \cite{BW93}, and quantum computation
\cite{EJ96,BDMT98}. A state of a composite quantum system is called
 ``entangled" if it can not be represented as a mixture of factorizable
pure states. Otherwise, the state is called separable. The above definition
is physically meaningful because entangled states (unlike separable states)
cannot be prepared locally by acting on each subsystem individually
 \cite{P93}. Nowadays there is general consensus on the fact that the
 phenomenon of entanglement is one of the most fundamental and
non-classical features exhibited by quantum systems \cite{LPS98}.

If one has enough information it is possible to determine the
amount of entanglement of a quantum system even if the available
information does not allow for a complete knowledge of the
system's state. An interesting example of this situation was
recently discussed by Sancho and Huelga, who studied the  minimal
experimental protocol required for determining the entanglement of
a two-qubits pure state from local measurements \cite{SH00}.
Another important result obtained by Sancho and Huelga is that the
knowledge of the expectation value of just one observable ({\it
local or not}) does not suffice to determine the entanglement of a
given unknown pure state of two particles \cite{SH00}. The case in
which the prior information is not sufficient for a complete
determination of the amount of entanglement was further examined
by Horodecki {\it et al.} \cite{HHH99}. These authors did not
restrict their analysis to pure states. They assumed that the
available information consists of the mean values of a given set
of observables $\hat A_i$. Jaynes' maximum entropy (MaxEnt)
principle \cite{B91,BAD96} provides a general inference scheme to
treat this kind of situations. According to Jaynes' principle, one
must choose the state yielding the least unbiased description of
the system  compatible with the available data. That state is
provided by the statistical operator  $\hat \rho_{ME} $ that
maximizes the von Neumann entropy $S \, = \, -Tr (\hat \rho \, \ln
\hat \rho)$ subject to the constraints imposed by normalization
and the expectation values $\langle \hat A_i \rangle \,= \, Tr
(\hat \rho \hat A_i) $ of the relevant observables $\hat A_i$.

Even though Jaynes' principle does provide a very satisfactory
answer in many situations \cite{B91,BAD96}, Horodecki {\it et al.}
\cite{HHH99} showed that the straightforward application of Jaynes'
prescription in its usual form is not always an appropriate strategy
for dealing with entangled states. It was shown in \cite{HHH99} that
the standard implementation of Jaynes' principle may create ``fake"
entanglement. For example, the MaxEnt density matrix may correspond
to an entangled state even if there exist separable states compatible
with the prior information. Since quantum {\it entanglement} is, in
many cases, the basic resource needed when processing quantum
information \cite{HHH99}, statistical inference procedures that
overestimate the amount of available entanglement should be handle
with care. Furthermore, it is  well-known that local operations and
classical communication (LOCC) can never increase the amount of
entanglement between remote systems, but they can make it
decrease \cite{LPS98}. As a consequence, one should often bet on the
decrease of entanglement and not be very ``optimistic" when estimating
the available amount of this resource. The above considerations
suggests that, in order to deal with some situations involving
entanglement, the usual form of Jaynes' prescription needs to be
modified or supplemented in an appropriate way. Various such schemes
have been proposed. Horodecki {\it et al.} \cite{HHH99} proposed a
combined strategy based on a constrained minimization of entanglement
followed by a maximization of the von Neumann entropy. Alternatively,
Abe and Rajagopal \cite{AR99} explored the possibility of inferring
entangled states by recourse to a variational principle based
on non-extensive information measures.

Up to now, all the work done in connection with
Horodecki's problem of fake inferred entanglement
focused on that particular case in which the prior information
is given by the mean value of the Bell operator
\cite{HHH99,RPPC00,R99,AR99}. The main purpose of
the present effort is to explore what happens when
the available prior information consists of the
expectation value of operators exhibiting a more general
form. Particular attention is going to be paid to operators
non diagonal in the Bell basis. We are going to show that
the prescription proposed in \cite{R99} for
solving the problem of fake entanglement is not
universally applicable. We will show that there
exist operators, both diagonal and non diagonal
in the Bell basis, for which the aforementioned
prescription fails.

The paper is organized as follows. In section II we revisit, from a different
point of view than the one employed in references \cite{HHH99,R99,AR99}, the
problem of ``fake entanglement" arising when a quantum state is inferred on
the basis of partial information related to the Bell observable. The inference
of entangled states from prior information associated with observables non
diagonal in the Bell basis is considered in sections III and IV. Finally,
some conclusions are drawn in section V.

\section{The Expectation Values of the Bell Observable and Its
Square as Input Information}

Following Horodecki {\it et al.} \cite{HHH99} let us assume
that the prior (input) information is given by the expectation value
$b$ of the Bell-CHSH observable \cite{CHSH69}

\begin{equation} \label{campana}
\hat B = \sqrt{2} \, \Bigl( \sigma_x\otimes \sigma_x+\sigma_z\otimes \sigma_z \Bigr)
= 2\sqrt{2} \, \Bigl(|\Phi^+\ra\la\Phi^+|-|\Psi^-\ra\la\Psi^-| \Bigr)\,
\end{equation}

\noindent
which is defined in terms of the components of the well-known Bell basis,

\begin{eqnarray} \label{babel}
|\Phi^\mp \ra &=& \frac{1}{\sqrt{2}} \, \Bigl(| 11 \ra\mp | 00 \ra \Bigr), \cr
|\Psi^\pm \ra &=& \frac{1}{\sqrt{2}} \, \Bigl(| 10 \ra\pm | 01 \ra \Bigr).
\end{eqnarray}

\noindent The Bell observable is {\it nonlocal}. In order to
measure the Bell observable one can not rely just upon local
operations and classical communication between the parts (that is,
LOCC operations). It can not be measured without interchange of
{\it quantum information} between the observers \cite{HHH99}.

\noindent
The MaxEnt state obtained by recourse to the standard prescription, when
the sole available information is given by $b=\langle \hat B \rangle $,
is described by the density matrix \cite{HHH99}

\begin{eqnarray} \label{rhoj}
\hat \rho_{ME}(b) &=&
\frac{1}{4}\Bigg[ \left(1+\frac{b}{\sqrt{2}}+\frac{b^2}{8}\right)
|\Phi^+\rangle \langle \Phi^+|
+ \left(1-\frac{b}{\sqrt{2}}+\frac{b^2}{8}\right)
|\Psi^-\rangle \langle \Psi^-|   \nonumber \\
& & + \left(1-\frac{b^2}{8}\right)
\Bigl( |\Psi^+\rangle \langle \Psi^+| +
|\Phi^-\rangle \langle \Phi^-| \Bigr) \Bigg].
\end{eqnarray}

Rajagopal \cite{R99} and Abe and Rajagopal \cite{AR99} showed that the inclusion
of $\sigma^2 = \langle \hat B^2 \rangle$ within the input data set entails important
consequences for the inference of entangled states. The main idea of Rajagopal's
proposal \cite{R99} is to consider the density matrix $\hat \rho_{MS} $ obtained by considering
both mean values $b=\langle \hat B \rangle$ and  $\sigma^2=\langle \hat B^2 \rangle $
as constraints in the MaxEnt prescription, and assuming that the mean value of
$\hat B^2$ adopts the minimum value compatible with the given value of $b$. Rajagopal
proved that $\hat \rho_{MS}$ is {\it separable} if and only if $b <\sqrt{2}$. The
method employed by Rajagopal to characterize the states $\hat \rho_{MS}$ of
minimum-$\sigma^2$ rests heavily on the particular form of the Bell operator. A
different approach is needed if one wants to implement Rajagopal's inference scheme
when the input information consists of the mean value of more general observables.
It is convenient now to briefly revisit the example corresponding to the Bell observable
in order to (i) illustrate the viewpoint that we are going to adopt when considering
more general situations, and (ii) to clarify the relationships between the results
we are going to report in this paper and those previously discussed in the literature.

The operators $\hat B$ and $\hat B^2$ verify the relations

\ben \label{b2b1}
\hat B^2 \, &=& \, 16 |\Phi^+ \rangle \langle \Phi^+ | \, - \, 2\sqrt{2} \hat B \cr
            &=& \, 16 |\Psi^- \rangle \langle \Psi^- | \, + \, 2\sqrt{2} \hat B.
\een

\noindent
It is easy to see, computing the trace of the above equations, that

\be \label{ineq1}
\sigma^2 \, \ge \, 2 \sqrt{2} \,\, |b|,
\ee

\noindent
and, consequently, the minimum value of $\sigma^2$  compatible
with a given value of $b$ is

\be \label{minimobdo}
\sigma^2 = 2 \sqrt{2} |b|.
\ee

\noindent From
  the trace of equation (\ref{b2b1}) it also transpires that  density
matrices with the minimum value of $\sigma^2$ compatible with a given value
of $b$ comply  with

\ben \label{sigmins}
\langle \Phi^+ |\hat \rho | \Phi^+ \rangle \, &=&  \, 0
\,\,\,\,\,\,\,\,\,\,\,\,\,\, ({\rm if} \,\,\, b<0) \cr
\langle \Psi^- |\hat \rho | \Psi^- \rangle \, &=&  \, 0
\,\,\,\,\,\,\,\,\,\,\,\,\,\, ({\rm if} \,\,\, b>0).
\een

\noindent
This means that a state complying with the  minimum uncertainty requirement belongs
to the three dimensional subspace spanned by the vectors
$\{|\Psi^+ \rangle, |\Psi^- \rangle, |\Phi^- \rangle \}$ ($b<0$),
or by the vectors
$\{|\Psi^+ \rangle, |\Phi^+ \rangle, |\Phi^- \rangle \}$ ($b>0$).
For the density matrices defined within this subspaces we have

 \ben \label{muc}
 b \, &=& \, - 2\sqrt{2} \, \langle   \Psi^- |\hat \rho| \Psi^- \rangle
 \,\,\,\,\,\,\,\,\,\,\,\,\,\,\,\,\,\, ({\rm if}\,\,\, b<0) \cr
 b \, &=& \,\,\,\,\,\, 2\sqrt{2} \, \langle   \Phi^+ |\hat \rho| \Phi^+ \rangle
 \,\,\,\,\,\,\,\,\,\,\,\,\,\,\,\,\,\, ({\rm if}\,\,\, b>0).
 \een

 \noindent
 The matrices provided by Rajagopal's scheme are

 \ben \label{rajarho}
 \hat \rho_{MS} \, &=& \, \frac{-b}{2\sqrt{2}}
 |\Psi^- \rangle \langle \Psi^- | \, + \, \frac{1}{2} \left(1+\frac{b}{2\sqrt{2}} \right)
 \left[ |\Psi^+ \rangle \langle \Psi^+ | + |\Phi^- \rangle \langle \Phi^- | \right]
 \,\,\,\,\,\,\,\,\,\,\,\,\,\,\,\,\,\, ({\rm if}\,\,\, b<0) \cr
 \hat \rho_{MS} \, &=& \, \frac{b}{2\sqrt{2}}
 |\Phi^+ \rangle \langle \Phi^+ | \, + \, \frac{1}{2} \left(1-\frac{b}{2\sqrt{2}} \right)
 \left[ |\Psi^+ \rangle \langle \Psi^+ | + |\Phi^- \rangle \langle \Phi^- | \right]
 \,\,\,\,\,\,\,\,\,\,\,\,\,\,\,\,\,\, ({\rm if}\,\,\, b>0)
 \een

\noindent
States that are diagonal in the Bell basis (\ref{babel}) are separable if and
only if they have no eigenvalue larger than $1/2$ \cite{HHH99}. Hence, it follows
from equation (\ref{rajarho}) that the states $\hat \rho_{MS}$ are separable if
and only if $|b| < \sqrt{2}$.

Let us now consider general minimum uncertainty states (that is, states
$\hat \rho$ verifying (\ref{minimobdo}) but not necessarily of the MaxEnt
form). Expressing the matrix elements of $\hat \rho$ in the Bell basis
(\ref{babel}), let us equate all the nondiagonal elements to zero and leave
unchanged the diagonal ones. The new density matrix $\hat \rho_D$ thus
obtained has always less entanglement than the original $\hat \rho$ \cite{HHH99}.
If the original $\hat \rho$ is such that $ b \, > \, \sqrt{2}$, then the
matrix $\hat \rho_D$ (which is diagonal in the Bell basis) will have one
eigenvalue greater than $1/2$ (see equation (\ref{muc})). Thus, $\hat \rho_D$ is
entangled and so is $\hat \rho$. Summing up, there is no separable density
matrix complying with the minimum-$\sigma^2$ condition (\ref{minimobdo}) and
having $b \, > \, \sqrt{2}$. This means that, for $b \, > \, \sqrt{2}$, the
inference scheme proposed by Rajagopal does not produce ``fake" inferred
entanglement. {\it At least when the input data is related to the Bell observable}
(\ref{campana}), {\it Rajagopal's prescription does not lead to an entangled inferred
state $\hat \rho_{MS}$ if there are separable states compatible with the constraints
$b$ and $\sigma^2$}. This is the main result obtained by Rajagopal \cite{R99,AR99},
although he arrived to it by recourse to a different line of reasoning.

 Quantitative measures of entanglement constitute interesting tools for
studying the entanglement-related properties exhibited by the
standard MaxEnt scheme and other statistical inference methods.
Notice that both Horodecki's and Rajagopal's discussions of the
problem of fake inferred entanglement only distinguish between
separable and entangled states. No degree of entanglement is
thereby ascertained. However, as it is well-known, entangled
states differ in the amount of entanglement they have. A
quantitative measure of entanglement enables us to compare the
degree of entanglement of both (i) the {\it inferred} quantum
state $\hat \rho_{inferred} $ yielded by an inference scheme when
only partial information is available about the {\it ``true"}
state $\hat \rho_{true}$ of the system and (ii) the entanglement
of $\hat \rho_{true}$. When both states $\hat \rho_{inferred}$ and
$\hat \rho_{true}$ are entangled, we would like to know the amount
of entanglement that each of these statistical operators carries
with it. A physically motivated measure of entanglement is
provided by the entanglement of formation $E[\hat \rho]$
\cite{BDSW96}. This measure quantifies the resources needed to
create a given entangled state $\hat \rho$. As explained in
references \cite{LPS98,BDSW96}, $E[\hat \rho]$ is equal to the
asymptotic limit (for large $n$) of a certain quotient $m/n$. Here
$m$ is the number of singlet states needed to create $n$ copies of
the state $\hat \rho$ when the optimum procedure based on local
operations is employed. Obviously, the entanglement of formation
of a separable state is equal to zero, that is $E(\hat
\rho_{sep.})=0$. For the particular case of two-qubits states,
Wootters obtained an explicit expression for $E[\hat \rho]$ in
terms of the density matrix $\hat \rho$ \cite{WO98}. Wootters'
formula reads \cite{WO98}

\be
E[\hat \rho] \, = \, h\left( \frac{1+\sqrt{1-C^2}}{2}\right),
\ee

\noindent
where

\be
h(x) \, = \, -x \log_2 x \, - \, (1-x)\log_2(1-x),
\ee

\noindent
and $C$ stands for the so-called {\it concurrence} of the two-qubits state
$\hat \rho$. The concurrence is given by

\be
C \, = \, max(0,\lambda_1-\lambda_2-\lambda_3-\lambda_4),
\ee

\noindent
$\lambda_i, \,\,\, (i=1, \ldots 4)$ being the square roots, in decreasing order,
of the eigenvalues of the matrix $\hat \rho \tilde \rho$, with

\be \label{rhotil}
\tilde \rho \, = \, (\sigma_y \otimes \sigma_y) \rho^{*} (\sigma_y \otimes \sigma_y).
\ee

\noindent The above expression is to be evaluated by recourse to
the matrix elements of $\hat \rho$ computed  with respect to the
product basis.

Fig. 1 depicts the entanglement of formation as a function of the input
 data $b$ (for $b>0$).
Two types of inferred density matrix are used to compute the
entanglement of formation, namely, (i) the density matrix $\hat
\rho_{ME}$ yielded by the standard MaxEnt procedure (upper solid
line) and (ii) the density matrix $\hat \rho_{MS}$ provided by
Rajagopal's minimum-$\sigma^2$ scheme (lower solid line).

Let us suppose that the ``true" state of the system is described by
a density matrix of the form

\be \label{alfalfa}
\hat \rho_T(\alpha)  =
\left(\frac{b}{2\sqrt{2}}+\alpha \right)|\Phi^+ \rangle \langle \Phi^+ |
 +  \alpha |\Psi^- \rangle \langle \Psi^- |  +
\frac{1}{2}\left(1 - \frac{b}{2\sqrt{2}} - 2\alpha \right)
\Bigl( |\Phi^- \rangle \langle \Phi^- | +
 |\Psi^+ \rangle \langle \Psi^+ | \Bigr).
\ee

\noindent The (``true") density matrices belonging to the above
family are characterized by a parameter $\alpha$ and verify
$Tr(\hat \rho_T \hat B) = b$. We assume that the only knowledge we
have about $\hat \rho_T$ is given by the mean value $b$. From this
piece of data we can determine the inferred matrices $\hat
\rho_{ME}$ and $\hat \rho_{MS}$ provided, respectively, by the
standard MaxEnt and Rajagopal's precriptions. In the inset of Fig.
1 we can see, together  with the entanglement of formation of both
$\hat \rho_{ME}$ and $\hat \rho_{MS}$, the behaviour (as a
function of $b$) of the entanglement of formation $E[\hat
\rho_T(\alpha)]$, i.e., that of the ``true" state.

We believe that the ($b,E(b)$)-plane depicted in Fig. 1,
representing input information $b$ versus the inferred
entanglement $E(b)$, constitutes a useful device for visualizing
the entanglement-related properties of an inference scheme. In
Fig. 1 we can compare how both the standard MaxEnt scheme, and the
one advanced by Rajagopal,  behave in the ($b,E(b)$)-plane. The
most noteworthy feature of Fig. 1 is that (when the input
information is related to the Bell observable) the results
obtained using the usual MaxEnt method do not seem to differ too
much from those obtained using Rajagopal's prescription.

\section{Input information associated with observables non diagonal in the
Bell basis}

As already mentioned, both Horodecki and Rajagopal treatments of the
problem of fake inferred entanglement focused on the case of prior
knowledge related to the Bell observable. We want to explore
here to what extent the conclusions reached by those researchers
are valid when the available prior information consists on the
expectation values of more general observables. In particular,
we want to explore what happens when observables non diagonal in
the Bell basis are considered. As we shall presently see, an interesting
example illustrating new aspects of the phenomenon of fake entanglement
is provided by the quantum observable associated with the hermitian
operator

\be \label{boperat}
\hat A \, = \, \kappa \Bigl(|1\rangle \langle 1| \, + \,
 |3\rangle \langle 3|\Bigr)
\, + \, \lambda |2\rangle \langle 2|,
\ee

\noindent
where $\kappa$ and $\lambda$ are real parameters such that

\ben \label{kala}
\kappa \ge 0 \ge \lambda ,
\een

\noindent
and whose eigenvectors
$|i\rangle\,\, (i=1,\ldots 4)$ are

\ben \label{basis}
|1\rangle \, &=& \, \frac{1}{\sqrt{2}} \, \Bigl( |11\rangle \, + \, |00\rangle  \Bigr), \cr
|2\rangle \, &=& \, \frac{1}{\sqrt{2}} \, \Bigl( |11\rangle \, - \, |00\rangle  \Bigr), \cr
|3\rangle \, &=& \, |01 \rangle, \cr
|4\rangle \, &=& \, |10 \rangle.
\een

\noindent {\it It is clear that $\hat A$ is non diagonal in the
Bell basis}. The observable $\hat A$ is nonlocal. It cannot be
measured without interchange of {\it quantum information} between
the observers. Consequently, and as far as its nonlocality
properties are concerned, the observable $\hat A$ has the same
status as the Bell observable considered by Horodecki
\cite{HHH99}, Rajagopal \cite{R99}, and Abe and Rajagopal
\cite{AR99}. Sancho and Huelga \cite{SH00} recently proved that
the  knowledge of the expectation value of just one observable
(even if the observable is nonlocal) is not enough to determine
completely the amount of entanglement of a given, unknown,
bipartite pure state. This important result immediately raises the
question of how reliably can the entanglement of an unknown
quantum state be inferred from the sole knowledge of the mean
value of a nonlocal observable. We are going to explore here some
aspects of this question, mainly in connection with the problem of
fake inferred entanglement. Let us suppose that we know the
expectation value $a$ of $\hat A$, given by

\be \label{valmeb}
a \, = \, Tr (\hat \rho \hat A) \, = \, \kappa
\Bigr( \langle 1 |\hat \rho | 1 \rangle + \langle 3 |\hat \rho | 3 \rangle \Bigl) \, + \,
\lambda \langle 2 |\hat \rho | 2 \rangle.
\ee

\noindent
Following the proposal first advanced in \cite{R99} (see also
\cite{RPPC00,AR99}) we
are going to incorporate a new constraint associated with the expectation
value of

\be \label{boperatsqu}
\hat A^2 \, = \, \kappa^2  \Bigl(|1\rangle \langle 1| \, + \, |3\rangle \langle 3|\Bigr)
\, + \, \lambda^2  |2\rangle \langle 2|,
\ee

\noindent
which is

\be \label{velmesig}
\sigma^2 \, = \, Tr (\hat \rho \hat A^2) \, = \, \kappa^2
\Bigr( \langle 1 |\hat \rho | 1 \rangle + \langle 3 |\hat \rho | 3 \rangle \Bigl) \, + \,
\lambda^2 \langle 2 |\hat \rho | 2 \rangle.
\ee

\noindent
According to the strategy suggested in \cite{R99}, the problem of fake inferred
entanglement can be solved if in order to describe
our system  we adopt a density
matrix
$\hat \rho_{MS}$ complying with two requisites.
First, $\hat \rho_{MS}$ must have the MaxEnt
form corresponding to the constraints associated with the expectation
values of both
$\hat A$ and $\hat A^2$.
Secondly, the expectation value $\sigma^2 $ must adopt
the lowest value compatible with the given value of $a$. Notice that the
mean value
$a=\langle \hat A \rangle$ is the only independent input data. For the sake of
simplicity we are going to restrict our considerations to the case of positive
values of $\langle \hat A \rangle$.

The mean values of $\hat A$ and $\hat A^2 $ are related by

\be \label{sigmabe}
\sigma^2 \,= \, \kappa a \, + \, \lambda (\lambda - \kappa) \langle 2 |\hat \rho | 2 \rangle,
\ee

\noindent
which implies that those mixed states  characterized by
exhibiting  the minimum possible
$\sigma^2$-value  compatible
with a given $a>0$ must verify $\langle 2|\hat \rho | 2 \rangle = 0$.
Consequently, for those
states with minimum $\sigma^2$ we have

\be \label{sigmamin}
\sigma^2 \, = \, \kappa a.
\ee

When we have a single constraint corresponding to the mean value of $\hat A$,
the maximum entropy density matrix is

\be \label{romeia}
\hat \rho_{ME}^I \, = \, \frac{1}{Z} \exp(-\beta \hat A),
\ee

\noindent
where $\beta$ is a Lagrange multiplier and $Z=Tr (\exp(-\beta \hat A))$.
Alternatively, $\hat \rho^I_{ME}$ can be cast as

\be \label{romeib}
\hat \rho_{ME}^I \, = \, \frac{1}{1 + 2w + w^{\lambda/\kappa}}\,
\left[ w \Bigl(|1\rangle \langle 1| \, + \, |3\rangle \langle 3|\Bigr) +
w^{\lambda/\kappa}|2\rangle \langle 2| + |4\rangle \langle 4| \right],
\ee

\noindent
where $w = \exp(-\beta \kappa)$ verifies

\be
\frac{a}{\kappa} \, = \,
\frac{2w + (\lambda/\kappa) w^{\lambda/\kappa}}{1 + 2w + w^{\lambda/\kappa}}.
\ee

\noindent
The maximum entropy statistical operator associated with
the expectation values $a$ and $\sigma^2$ as input information
is

\be \label{roMaxEnta}
\hat \rho_{ME}^{II} \, = \, \frac{1}{Z} \exp(-\beta \hat A -\gamma \hat A^2),
\ee

\noindent
where $\beta $ and $\gamma$ are appropriate Lagrange multipliers and the
partition function $Z$ is given by

\be \label{z2}
Z=Tr (\exp(-\beta \hat A -\gamma \hat A^2)).
\ee

\noindent
The matrix $\hat \rho^{II}_{ME}$ can be expressed explicitly in terms
of the input mean values $a$ and $\sigma^2$,

\be \label{roMaxEntb}
\hat \rho_{ME}^{II} \, = \,
\frac{1}{2} \frac{\sigma^2 - \lambda a}{\kappa(\kappa - \lambda)} \,
\Bigl(|1\rangle \langle 1| \, + \, |3\rangle \langle 3|\Bigr)
\, + \,
\frac{\kappa a - \sigma^2}{\lambda(\kappa - \lambda)} \,
|2\rangle \langle 2|
\, + \,
\frac{\sigma^2 - a (\kappa + \lambda) + \lambda \kappa}{\lambda \kappa} \,
|4\rangle \langle 4|.
\ee

\noindent
When the further requirement of
a minimum value for $\sigma^2 $ is imposed, the above MaxEnt
density matrix reduces to

\be \label{romasigmin}
\hat \rho_{MS} \, = \,
\frac{a}{2\kappa} \,
\Bigl(|1\rangle \langle 1| \, + \, |3\rangle \langle 3|\Bigr)
\, + \, \left( 1 - \frac{a}{\kappa} \right)\,
|4\rangle \langle 4|.
\ee

\noindent
Since we always have $\kappa \ge a$, the above matrix is positive semidefinite.

Now, in order to find out whether Rajagopal's prescription is
plagued with the problem of fake inferred entanglement (when
applied in connection with the observable $\hat A$), we need to
proceed according to what follows. First, we adopt  a form for the
``true" density matrix describing the system. Second, we assume
that the only available information about the true state consists
on the expectation value of $\hat A$. From this sole piece of data
we obtain, via the inference scheme we are studying, the inferred
density matrix. Finally, we compare the entanglement properties
associated with
 the original, true density matrix with the entanglement properties
exhibited by the inferred one. In particular, we can evaluate on
both matrices an appropriate  quantitative measure of
entanglement. In what follows we are going to assume that the true
state of the system is described by an statistical operator
belonging to the family of density matrices

\be \label{rosepar}
\hat \rho_{S} \, =
\, p |1\rangle \langle 1| \, + \, \alpha |3\rangle \langle 3|
\, + \, (1-p-\alpha ) |4\rangle \langle 4|,
\ee

\noindent
where $p$ and $\alpha $ are real positive parameters verifying

\ben \label{palfa}
0 &\le & p \le 1 \cr
0 &\le & \alpha \le 1-p.
\een

\noindent Notice that the ``true" density matrices (\ref{rosepar})
that we are trying to infer by recourse to different schemes are
not of the maximum entropy form, nor of the form associated with
any other statistical inference scheme. The expectation values of
$\hat A $ and $\hat A^2 $, evaluated on $\hat \rho_S$ are

\be \label{bsep}
a \, = \, p \kappa \, + \, \alpha \kappa,
\ee

\noindent
and

\be \label{sigsep}
\sigma^2 \, = \, p \kappa^2 \, + \, \alpha \kappa^2.
\ee

\noindent
Suppose we are given the expectation values $a$
and $\sigma^2 $ corresponding to a given state
belonging to the family (\ref{rosepar}) (notice
that, for this family of density matrices, the
mean values $a$ and $\sigma^2$ always verify the
minimum-$\sigma^2$ condition (\ref{sigmamin})). We can
take those mean values as input information and
generate the concomitant inferred density matrix.
That is, we can associate a MaxEnt state to each
member of (\ref{rosepar}). The performance of the
inference scheme can be studied by comparing the
entanglement properties of a member of the
parameterized family (\ref{rosepar}) with those of
the concomitant inferred state. As a first step we
are going to find out, by recourse to Peres'
separability criterion \cite{P96}, whether there
are separable states of the form (\ref{rosepar})
leading to entangled inferred states. Peres'
criterion is based on a partial transposition
transformation \cite{P96}. To be more specific,
let the density matrix elements (with respect to
a product basis) of a statistical operator
$\hat \rho$ be

\be  \label{peres1}
\rho_{m \mu, n \nu} \, = \, \langle m \mu |\hat \rho |n \nu \rangle,
\ee

\noindent
where Latin indices refer to the first subsystem and Greek indices to the
second one. The partial transpose $\hat \rho^{PT} $ of $\hat \rho$ is a
matrix whose elements are obtained by the partial transposition of the
elements of $\hat \rho$, i.e.,

\be \label{transpar}
\hat \rho^{PT}_{{m \mu, n \nu}} \, = \, \hat \rho_{n \mu, m \nu}.
\ee

\noindent
It can be shown that  $\hat \rho$ is separable if and
only if $\hat \rho^{PT}$ has no negative eigenvalues
\cite{HHH96}. If we apply the Peres' criterion to
the minimum-$\sigma^2$ MaxEnt density matrix $\hat
\rho_{MS}$ (Eq. \ref{romasigmin}) we find that
there is only one eigenvalue of the partial
transpose matrix that may adopt negative values.
This eigenvalue is

\be \label{negati}
\delta \, = \,
- \frac{a}{4\kappa} \, + \, \frac{1}{2} \, - \, \frac{1}{4}
\sqrt{\frac{a}{\kappa} \, \left(10 \frac{a}{\kappa} - 12 \right) + 4}.
\ee

\noindent
Hence, we have

\ben \label{peres} a/ \kappa \, &\le & 8/9 \, \Longleftrightarrow
\, \delta \ge 0,
 \cr a/ \kappa \,  &>&  8/9 \, \Longleftrightarrow \,
\delta   < 0  \een

\noindent Consequently, $\hat \rho_{MS}$ is separable if $a/\kappa
\le 8/9$ and entangled otherwise. Using the Peres' criterion we
can also determine just when the parameterized (true) density
matrix $\hat \rho_S$ is separable. For the considerations that
follow it will prove convenient to rewrite $\hat \rho_S$ in terms
of the expectation value $a= Tr(\hat \rho_S \hat A)$,

\be \label{rose} \hat \rho_S \, = \, \left(\frac{a}{\kappa} -
\alpha \right) |1\rangle \langle 1| \, + \, \alpha |3\rangle
\langle 3| \, + \, \left(1 - \frac{a}{\kappa} \right) |4\rangle
\langle 4|.
 \ee

\noindent
 It is important to stress that the above expression describes
 the same family of mixed states defined by equation
 (\ref{rosepar}). The states $\hat \rho_S$ associated with
 equation (\ref{rose}) still depend on two independent parameters,
 i.e., $\alpha$ and $a/\kappa$. Equation (\ref{rose}) is just a
 re-parameterization of the family (\ref{rosepar}) where, for the
 sake of convenience, we have chosen
 $a/\kappa=Tr(\hat \rho_S \hat A)/\kappa$
  as one of the two relevant parameters. The separability of
  $\hat \rho_S$ is determined by the quantity

\be
Q \, = \,\frac{1}{2} \, - \, \frac{a}{2\kappa} \, + \, \frac{\alpha}{2} \, - \, \frac{1}{2}
\sqrt{2 \left(\frac{a}{\kappa} \right)^2 - \frac{2a}{\kappa} + 1 -2\alpha + 2\alpha^2}.
\ee

\noindent The statistical operator $\hat \rho_S$ is separable if
$Q\ge 0$ and entangled otherwise. The boundaries (in the plane
$(\alpha,a)$) between the separability and the entangled regions
corresponding to (i) the density operators $\hat \rho_S$, (ii) the
standard MaxEnt statistical operators $\hat \rho_{ME}^I$, and
(iii) the minimum-$\sigma^2$ MaxEnt density matrices $\hat
\rho_{MS}$, are depicted in Fig. 2, where we take $\kappa=1$ and
$\lambda=-1$. Notice that only those points with $\alpha <a$ are
physically meaningful, since $(\alpha,a)$ pairs not complaining
with that inequality lead to a matrix $\hat \rho_S$ with one
negative eigenvalue. Figure 2 is to be interpreted as follows.
There are three density matrices associated with each point in the
plane $(\alpha,a)$:

\begin{itemize}
\item{(i) The (``true") $\hat \rho_S$ matrix given by
the expression (\ref{rose})}.
\item{(ii) The (inferred) density matrix $\hat \rho_{ME}^I$}, of the
standard MaxEnt form (\ref{romeia}-\ref{romeib}).
\item{(iii) The (inferred) density matrix  $\hat \rho_{MS}$ of  the
minimum-$\sigma^2$ MaxEnt form
(\ref{romasigmin})}.
\end{itemize}

\noindent For all the three aforementioned density matrices
 the  expectation value of $\hat A$ is $a$,
 (that is,
$a = Tr(\hat \rho_{MS} \hat A)=Tr(\hat \rho_{ME}^I)=Tr(\hat
\rho_{S} \hat A)$). The density matrix $\hat \rho_{MS}$ is the one
yielded by Rajagopal's prescription if one tries to infer $\hat
\rho_S$ from the sole knowledge of the expectation value $a = Tr
(\hat \rho_S \hat A) $. The standard MaxEnt procedure, instead,
would lead to $\hat \rho_{ME}^I$. Using the Peres' criterium we
can determine when the inferred density matrix $\hat \rho_{ME}^I$
is entangled. For $\kappa=1$ and $\lambda=-1$ we found that $\hat
\rho_{ME}^I $ is separable when $a\le 0.8564$ and entangled
otherwise. The lines $l$ and $m$ in Fig. 2 corresponds to
$a=0.8564 $ and $a=8/9$, respectively. The curve $n$ represents
the equation $Q(a,\alpha)=0$. The density matrices $\hat
\rho_{ME}$ ($\hat \rho_{MS}$) are entangled for points
$(a,\alpha)$ lying above the line $l$ ($m$) and separable
otherwise. On the other hand, the matrices $\hat \rho_S $ are
separable when $(a,\alpha)$ lies below the curve $n$ and entangled
if $(a,\alpha)$ lies above $n$. Of particular interest are the
regions I and II. {\it In region I the (``true") density matrix to
be inferred, $\hat \rho_S$, is separable, while the associated
(``inferred") matrix $\hat \rho_{MS}$, provided by Rajagopal's
inference scheme, is not}. In region II things are quite
different: the inference scheme provides a separable statistical
operator $\hat \rho_{MS}$ while the matrix to be inferred, $\hat
\rho_S$ is entangled. It is clear that {\it the maximum entropy
minimum-$\sigma^2$ inference procedure advanced by Rajagopal}
\cite{R99,AR99} {\it generates fake entanglement when applied to
states $\hat \rho_S$ associated with points $(a,\alpha)$ belonging
to region I }. Contrary to previous evidence obtained when the
Bell's observable mean value is taken as the prior information
\cite{RPPC00,R99,AR99}, we must conclude that the MaxEnt
minimum-$\sigma^2$ scheme does not provide a general solution to
the problem of fake entanglement.

 The comparison of the amount of entanglement of formation exhibited by
the states
 $\hat \rho_S$ and $\hat \rho_{MS}$ enables us to study the problem of fake
inferred
 entanglement in a quantitative way. The curves depicted in Fig. 3
display the
 behaviour of $E[\hat \rho_{MS}]$ and $E[\hat \rho_S]$ as a function of
the mean value
 $a$ of the observable $\hat A$ (again, with $\kappa=1$ and $\lambda=-1$).
The upper solid line
corresponds to $E[\hat \rho_{ME}^{I}]$, the lower solid line to $E[\hat
\rho_{MS}]$,
and the dashed and dot-dashed lines to $E[\hat \rho_S]$, for different values of the
parameter $\alpha$.
The results exhibited in
Fig. 3 illustrate how, for each given value of the input data $a=Tr(\hat \rho
\hat A)$, the
entanglement of formation $E$ of
 the density operators yielded by both the standard MaxEnt method
($\hat \rho_{ME}^{I}$) and Rajagopal's scheme ($\hat \rho_{MS}$) compare
with the  entanglement
of formation
of the state to be inferred ($\hat \rho_{S}$). It is clear from Fig. 3
that, with regards to
the behaviour of the inferred amount of entanglement as a function of the
input information
(at least when this input data consists of $\langle \hat A \rangle$),
the prescription advanced by Rajagopal does not appreciably
differ  from the
standard MaxEnt result.
In particular, both prescriptions tend to yield the same results
in the limit
$a\rightarrow 1$.

 Notice that the MaxEnt minimum-$\sigma^2$ matrix $\hat \rho_{MS}$ does
not depend upon
the value of $-(\lambda/\kappa)$, unlike what happens with the standard
MaxEnt matrix
$\hat \rho_{ME}^{I}$. This dependence upon $-(\lambda/\kappa)$ is depicted
in Fig. 4, where
we can
appreciate
 the behaviour of the entanglement of formation $E[\hat \rho]$ as
a function
of $(a/\kappa)$ corresponding to
i) the density operators $\hat \rho_{MS}$
(lower solid line)
and ii)
 the MaxEnt density matrices $\hat \rho_{ME}^{I}$ associated with
different values of the ratio $-(\lambda/\kappa)$
(dashed lines). The upper solid line in Fig. 4
corresponds to the particular case
$-(\lambda/\kappa)=1$. The MaxEnt density matrices
$\hat \rho_{ME}^{I}$ are entangled for values of
$a$ greater than a critical value $a_c$ depending
on $-(\lambda/\kappa)$. The behaviour of
$(a_c/\kappa)$ as a function of
$-(\lambda/\kappa)$ is depicted in the inset of
Fig. 4.

\section{Prior Information Associated with More General Observables}

In this section we are going to assume that the prior information
is given by the expectation value of an observable of the form

\be \label{observad}
\hat D \, = \, |1\rangle \langle 1| \, + \,  \alpha_1 |2\rangle \langle 2| \, + \,
\alpha_2 |3\rangle \langle 3|,
\ee

\noindent
with eigenvectors

\ben \label{titabas}
|1\ra \, &=& \, |\Phi^{+} \ra \cr
|2\ra \, &=& \, |\Phi^{-} \ra \cr
|3\ra \, &=& \, \sin \theta |10\ra + \cos \theta |01\ra \cr
|4\ra \, &=& \, \cos \theta |10\ra - \sin \theta |01\ra,
\een

\noindent
and eigenvalues ${0,1,\alpha_1,\alpha_2}$, such that

\be \label{desigualf}
\alpha_2 > \alpha_1 > 1.
\ee

\noindent The operator $\hat D$ can not be measured using
 only LOCC operations. In this respect it behaves like both
 the Bell observable and the observable $\hat A$ introduced
 in the preceding section.

 The mean values $d=\la \hat D \ra$ and $\sigma^2=\la \hat D^2 \ra$ are
 related by

\be \label{sigmad}
\sigma^2 - d \, = \,  \alpha_1(\alpha_1-1) \la 2| \hat \rho |2 \ra \, + \,
\alpha_2 (\alpha_2-1) \la 3| \hat \rho |3 \ra.
\ee

\noindent
In order to apply the inference method advanced by
Rajagopal we need first to determine the form
adopted by the statistical operators $\hat \rho$
characterized by the minimum possible value of
$\sigma^2$ compatible with a given value of $d$.
As we will presently see, the particular form
exhibited by the minimum-$\sigma^2$ density
matrices depends on the value of the constraint
$d$. It is clear from (\ref{observad})
and (\ref{desigualf}) that $0\le
d \le \alpha_2$. The minimum-$\sigma^2$ matrices
adopt three different forms associated,
respectively, with $d$-values belonging to the
intervals $[0,1]$, $[1,\alpha_1]$, and
$[\alpha_1,\alpha_2]$. It follows from
(\ref{sigmad}) that

\be \label{optirho1}
0 \le d \le 1 \Rightarrow \hat \rho_{MS} =
d |1 \ra \la 1| + (1-d) |4 \ra \la 4|.
\ee

\noindent
In order to analyze the case corresponding to $d\in[1,\alpha_1]$ it will
prove convenient to introduce the definitions

\ben \label{peses}
p   \, &=& \, \la 1 |\hat \rho |1 \ra, \cr
S   \, &=& \, \la 2 |\hat \rho |2 \ra \, + \, \la 3 |\hat \rho |3 \ra, \cr
s_1 \, &=& \, \la 2 |\hat \rho |2 \ra/S, \cr
s_2 \, &=& \, \la 3 |\hat \rho |3 \ra/S.
\een

\noindent
All the above quantities belong to the interval $[0,1]$. Furthermore, we have
$s_1+s_2=1$ and $0\le p+S \le 1$. The expectation value of $\hat D$ is given
by

\be \label{spd}
d \, = \, Tr(\hat \rho \hat D) \, = \, p \, + \,
S \Bigl(s_1 \alpha_1 + s_2\alpha_2 \Bigr),
\ee

\noindent
and the minimization of $\sigma^2$ is equivalent to
finding the minimum value of the quantity

\be \label{mina}
M \, = \,
S \Bigl(s_1 \alpha_1(\alpha_1-1) + s_2\alpha_2(\alpha_2-1) \Bigr).
\ee

\noindent
The variables $p$, $S$, and $s_1$ verify

\be \label{pemases}
0 \le p+S \, = \, d \, - \,
S\Bigl( s_1(\alpha_1-1) + (1-s_1)(\alpha_2-1)\Bigr) \le 1.
\ee

\noindent
Notice that once a particular value of $d$ is
fixed the parameters $p$, $S$, and $s_1$ are no
longer independent quantities: they are related by
(\ref{spd}) (which is equivalent to the equality
relation in (\ref{pemases})). Regarding $S$ and
$s_1$ as independent quantities, the optimization
problem we have to solve is to find the pair of
numbers $(S,s_1)$ belonging to $[0,1]$ that,
complying with the inequalities in
(\ref{pemases}), make $M$ a minimum. If we are
given a pair $(S,s_1)$ satisfying the
aforementioned requisites, it is clear that we can
decrease $S$ until the last inequality in
(\ref{pemases}) becomes an equality. Hence, the
optimum $(S,s_1)$ must verify

\be \label{eseig}
S \, = \, \frac{d-1}{s_1(\alpha_1-1) + (1-s_1)(\alpha_2-1)},
\ee

\noindent
and $M$ can be rewritten as a function of the sole variable
$s_1$ (remember that $s_2=1-s_1$)

\be \label{emeig}
M \, = \, (d-1) \,
\frac{s_1 \alpha_1(\alpha_1-1) + s_2\alpha_2(\alpha_2-1) }
{s_1(\alpha_1-1) + (1-s_1)(\alpha_2-1)}.
\ee

\noindent
Notice that the expression $(\ref{eseig})$ determines a value of $S$
that, for any value of $s_1\in [0,1]$, belongs to the interval $[0,1]$.
Introducing now the quantities

\ben \label{tete}
t_1 \, &=& \, s_1(\alpha_1-1)/[s_1(\alpha_1-1) + (1-s_1)(\alpha_2-1)], \cr
t_2 \, &=& \, (1-s_1)(\alpha_2-1)/[s_1(\alpha_1-1) + (1-s_1)(\alpha_2-1)],
\een

\noindent
the function $M$ to be minimized can be cast under the guise

\be \label{emete}
M \, = \, (d-1)(t_1 \alpha_1 + t_2 \alpha_2),
\ee

\noindent
which clearly adopts its minimum value when $t_1=1$ and $t_2=0$. That is,
 the minimum obtains when $s_1=1$. Summing up, the minimum-$\sigma^2$
density matrix compatible with a given value of $d\in [1, \alpha_1]$
corresponds to

\ben \label{esebalf}
S \, &=& \, \frac{d-1}{\alpha_1-1}, \cr
s_1 \, &=& \, 1.
\een

\noindent
The concomitant density operator reads

\be \label{optirho2}
\hat \rho_{MS} \, = \,
\left( \frac{\alpha_1-d}{\alpha_1-1} \right) |1\ra \la 1| \, + \,
\left( \frac{d-1}{\alpha_1-1} \right) |2\ra \la 2|.
\ee

\noindent
A similar reasoning can be applied in order to obtain $\hat \rho_{MS}$ when
$\alpha_1 \le d \le \alpha_2$. In this case, however, the variable $t_1$ in
equations (\ref{tete},\ref{emete}) can not reach the value $1$ because that
would imply  $S>1$ in (\ref{eseig}). Since the largest possible value of
$S$ is $1$, the optimum value of $t_1$ (and of $s_1$) is the one making
$S=1$ in (\ref{eseig}). This, in turn, implies that $p=0$. In this case
the minimum-$\sigma^2$ density matrix is

\be \label{optirho3}
\hat \rho_{MS} \, = \,
\left( \frac{\alpha_2-d}{\alpha_2-\alpha_1} \right) |2\ra \la 2| \, + \,
\left( \frac{d-\alpha_1}{\alpha_2-\alpha_1} \right) |3\ra \la 3|.
\ee

\noindent
An interesting feature of the minimum-$\sigma^2$ density matrix
associated with $\hat D$ is that, for this observable, the
requirement of minimizing $\langle \hat D^2 \rangle$ under the
constraint imposed by $\langle \hat D \rangle$ completely determines
the matrix $\hat \rho_{MS}$. That is, the maximum entropy
principle plays no role whatsoever
when implementing Rajagopal's prescription
for the observable $\hat D$. This seems to be a consequence of the
non degenerate character of the eigenvalues of $\hat D$.
  The entanglement of formation $E(\hat \rho_{MS})$ of the
minimum-$\sigma^2$ state, as a function of the
input data $d=\langle \hat D \rangle$, is compared
in Fig. 5, for different values of $\theta $, with
the entanglement of formation $E(\hat \rho_{ME})$
of the standard MaxEnt state

\be
\hat \rho_{ME} \, = \, \frac{1}{Z} \, \exp(-\beta \hat D),
\ee

\noindent
where $Z=Tr (\exp(-\beta \hat D))$. The most remarkable feature
of Fig. 5 is that, for extended ranges of $d$-values, the
minimum-$\sigma^2$ state is much more entangled than the standard
MaxEnt state. Hence, in this case $\hat \rho_{MS}$ is likely to
create a larger amount of fake inferred entanglement than the one
generated by $\hat \rho_{ME}$. As a matter of  fact, those values
of $d$ leading to a separable MaxEnt matrix $\hat \rho_{ME}$ and
to an entangled matrix $\hat \rho_{MS}$ provide explicit examples
of fake entanglement generated by Rajagopal's scheme, the standard
MaxEnt matrix itself describing the separable state compatible with
the input information. It is remarkable that this occurs even in the
case $\theta = \pi/4$, corresponding to input data associated with
an observable diagonal in the Bell basis.

Finally, notice that the study we have done in this Section can be
extended to the general case where the input information consists
of the expectation value of an arbitrary observable endowed with a
non degenerate spectra. Given an observable

\be \label{nodege}
\tilde D \, = \, \sum_{i=1}^4 \, d_i |i\rangle \langle i|,
\ee

\noindent
with $d_1<d_2<d_3<d_4$, let us consider the new observable

\be \label{nodegedos}
\hat D \, = \, \frac{\tilde D - d_1 \hat I}{d_2 - d_1}.
\ee

\noindent
It is clear that the operator $\hat D$ is of the form (\ref{observad}),
with $\alpha_1=(d_3-d_1)/(d_2-d_1)$ and $\alpha_2=(d_4-d_1)/(d_2-d_1)$,
and that the minimization of $\langle \hat D^2 \rangle$ for a
given value of $\langle \hat D \rangle$ is equivalent to the
minimization of $\langle \tilde D^2 \rangle$  for a given
value of $\langle \tilde D \rangle$.

\section{Conclusions}

As shown by Horodecki {\it et al.} \cite{HHH99},
the quantum state obtained by recourse to the
standard Maxent inference prescription may be an
entangled one even if there exist separable states
compatible with the prior data. This situation
constitutes a particularly clear instance of the
problem of ``fake" inferred entanglement. In order
to overcome this difficulty, but still within the
strictures of the standard MaxEnt perspective,
Rajagopal advanced an alternative approach to the
inference of entangled states \cite{R99}. His idea
is that of considering the maximum entropy state
consistent with both the mean value of the
observable $\hat A $ one is interested in and the
mean value of its square $\hat A^2$, adopting for
$\la \hat A^2 \ra$ the minimum value compatible
with a given value of $\la \hat A \ra$. In the
case of Horodecki's example (where the prior
information consists of the expectation value of
the Bell operator) Rajagopal's procedure yields a
separable state whenever there are separable
states compatible with the available data
\cite{R99}. This, together with other results
recently reported in the literature
\cite{RPPC00,AR99}, constituted evidence
supporting the idea that the minimum-$\sigma^2$
scheme may provide an appropriate and general way
of solving the problem of fake inferred
entanglement. However, all the aforementioned
evidence was based on the study of particular
examples in which the prior information was
related to the Bell-CHSH observable (diagonal in
the Bell basis). In order to find out to what
extent the minimum-$\sigma^2$ prescription
provides a reliable inference scheme of general
applicability, we have explored here its
performance when the prior information is related
to more general observables, emphasizing those
situations involving observables non diagonal in
the Bell basis. We have found explicit examples,
related to this kind of observables, in which the
minimum-$\sigma^2$ inference procedure leads to
entangled density matrices even if there exist
separable states compatible with the input data.
This means that the minimum-$\sigma^2$
prescription is not free from the fake
entanglement difficulty.

  There is no doubt that Jaynes'  MaxEnt principle has to
play an important role in any appropriate scheme
for the inference of entangled quantum states.
Indeed, one of the most remarkable features of
Jaynes' principle is its robustness: usually, when
it seems to fail, the real problem is not the
inadequacy of the MaxEnt principle itself, but
rather that some piece of relevant (prior)
information is not being taken into account. As
recently pointed out by Brun, Caves and Schack
\cite{BCS00}, the various inference schemes
recently advanced to solve the fake inferred
entanglement problem admit of an interpretation
within the strictures of Jaynes' approach. These
inference prescriptions may be regarded as
implementations of the MaxEnt principle in which
some extra prior information (that may not
consists just of the expectation values of some
observables) is assumed to be known. This is
certainly the case with Rajagopal's MaxEnt
minimum-$\sigma^2$ proposal, which assumes extra
information related to the square of the relevant
observable. However, the results reported here
show that this approach works only in very special
situations.

Besides enabling us to asses the usefulness of the minimum-$\sigma^2$
scheme, the present effort also sheds some new light on the
entanglement features exhibited by the standard MaxEnt principle
within contexts more general than those previously considered in
the literature \cite{HHH99,RPPC00,R99,AR99}.

\acknowledgments
This work was partially supported by the AECI Scientific Cooperation
Program, by the DGES grants PB98-0124 and SB97-26373862
(Spain), and by CONICET
(Argentine Agency).

\begin{figure}
\caption{
The entanglement of formation $E[\hat \rho]$, as a
function of i) the expectation value $b$ of the
Bell operator, ii) the MaxEnt density matrix $\hat
\rho_{ME}$ (Eq. (\ref{rhoj})) (upper solid line),
and iii) the minimum-$\sigma^2$ density matrix
$\hat \rho_{MS}$ (Eq. (\ref{rajarho})) (lower
solid line). The results corresponding to the
density matrix ansatz (\ref{alfalfa}) (dashed
lines) are shown in the inset.}
\label{f1}
\end{figure}

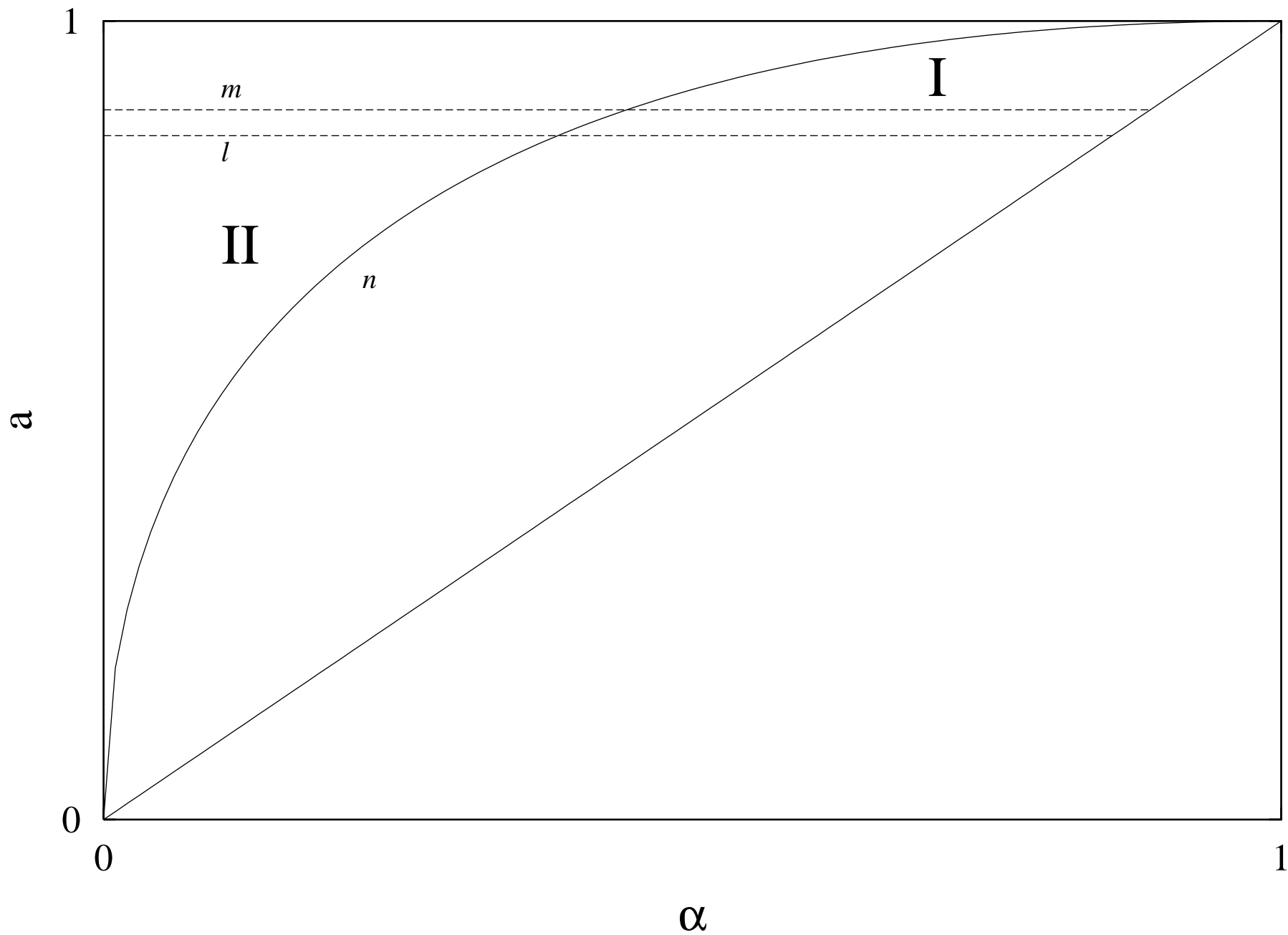
\begin{figure}
\caption{Boundaries between the regions corresponding
to separability and entanglement for states
described by the density matrices $\hat
\rho_{ME}^{I}$ (line $l$), $\hat \rho_{MS}$ (line
$m$), and $\hat \rho_S$ (line $n$). The
expressions for the matrices $\hat
\rho_{ME}^{I}$, $\hat \rho_{MS}$, and $\hat
\rho_S$ are given, respectively, by equations
(\ref{romeib}), (\ref{romasigmin}), and
(\ref{rose}).}
\label{f2}
\end{figure}

\begin{figure}
\caption{The entanglement of formation $E[\hat \rho]$ as a
function the expectation value of the observable $\hat A$ (Eq.
(\ref{boperat})) with $\kappa =1$ and $\lambda=-1$, corresponding
to $\hat \rho_{ME}^{I}$ (upper solid line), to $\hat \rho_{MS}$
(lower solid line), and to $\hat \rho_S$, for the values of
$\alpha$ indicated in the figure (dashed and dot-dashed lines).
The expressions for the matrices $\hat \rho_{ME}^{I}$, $\hat
\rho_{MS}$, and $\hat \rho_S$ are given, respectively, by
equations (\ref{romeib}), (\ref{romasigmin}), and (\ref{rose}).}
\label{f3}
\end{figure}

\begin{figure}
\caption{ The entanglement of formation $E[\hat \rho]$ as a
function of $(a/\kappa)$, where $a$ is the expectation value of
the observable $\hat A$ (Eq. (\ref{boperat})), corresponding to
$\hat \rho_{MS}$ (lower solid line) and to the MaxEnt density
matrices $\hat \rho_{ME}^{I}$ associated with different values of
the ratio $-(\lambda/\kappa)$ (dashed lines). The upper solid line
corresponds to the particular case $-(\lambda/\kappa)=1$. The
expressions for the matrices $\hat \rho_{ME}^{I}$ and $\hat
\rho_{MS}$ are given, respectively, by equations (\ref{romeib})
and (\ref{romasigmin}). The critical values $(a_c/\kappa)$ where
the matrices $\hat \rho_{ME}^{I}$ begin to be entangled are
depicted in the inset as a function of $-(\lambda/\kappa)$.}
\label{f4}
\end{figure}

\begin{figure}
\caption{The entanglement of formation as a function of
the expectation value of the observable $\hat D$ with
$\alpha_1=2$ and $\alpha_2=3$ (see
equations (\ref{observad}-\ref{desigualf})) evaluated,
for different values of $\theta$, on (i) the MaxEnt density
matrix $\exp(-\beta \hat D)/Tr(\exp(-\beta \hat D)$ (solid lines)
and (ii) the state exhibiting the minimum value of
$\langle \hat D^2 \rangle $ compatible with
$\langle \hat D \rangle $ (dashed lines). }
\label{f5}
\end{figure}


\begin{thebibliography}{}

\bibitem{HHH99} Horodecki R, Horodecki M and Horodecki P 1999
 Phys. Rev. A {\bf 59} 1799

\bibitem{BDADK97} Buzek V Drobny G Adam G Derka R and  Knight P L 1997
J. Mod. Opt. {\bf 44}  1607

\bibitem{RPPC00} Rigo A, Plastino A R, Plastino A, and Casas M
2000 Phys. Lett. A {\bf 270} 1

\bibitem{R99}  Rajagopal A K 1999 Phys. Rev. A {\bf 60} 4338

\bibitem{AR99} Abe S and  Rajagopal A K 1999 Phys. Rev. A {\bf 60} 3461

\bibitem{SH00}  Sancho J M G and Huelga S F 2000 Phys. Rev. A {\bf 61}
 1

\bibitem{BCS00} Brun  T A, Caves  C M and Schack  R {\it Entanglement
Purification of Unknown Quantum States}, quant-ph/0010038.

\bibitem{LPS98} Hoi-Kwong Lo, Popescu S and  Spiller T (Editors) 1998
{\it Introduction to Quantum Computation and Information}
(River Edge: World Scientific)

\bibitem{WC97}  Williams C P and Clearwater S H 1997 {\it Explorations in
Quantum Computing} (New York: Springer).

\bibitem{W98} Williams C P (Editor) 1998 {\it Quantum Computing and Quantum
Communications} (Berlin: Springer)

\bibitem{E91} Ekert A 1991 Phys. Rev. Lett. {\bf 67} 661

\bibitem{BBCJPW93} Bennett C H, Brassard G, Crepeau C, Jozsa R,
Peres A and  Wootters W K 1993 Phys. Rev. Lett. {\bf 70} 1895

\bibitem{BW93}Bennett C H and Wiesner S J 1993 Phys. Rev. Lett. {\bf 69}
 2881

\bibitem{EJ96} Ekert A and Jozsa R 1996 Rev. Mod. Phys. {\bf 68} 733

\bibitem{BDMT98}  Berman G P, Doolen G D, Mainieri R, Tsifrinovich V I
1998 {\it Introduction to Quantum Computers} (Singapore: World Scientific)

\bibitem{P93} Peres A 1993 {\it Quantum Theory: Concepts and Methods}
(Dordrecht: Kluwer)

\bibitem{B91} Balian R 1991 {\it From Microphysics to Macrophysics}
(Berlin: Springer)

\bibitem{BAD96} Buzek V, Adam G and Drobny G 1996 Ann. Phys. (N.Y.)
{\bf 245}  36

\bibitem{CHSH69}  Clauser J F,  Horne M A, Shimony A and Holt R A
1969 Phys. Rev. Lett. {\bf 23} 880

\bibitem{BDSW96}  Bennett C H, DiVicenzo D P, Smolin J and  Wootters W K
1996 Phys. Rev. A {\bf 54} 3824

\bibitem{WO98} Wootters  W K 1998 Phys. Rev. Lett. {\bf 80}  2245

\bibitem{PMPY97} Plastino  A R, Miller H G, Plastino A and Yen G D
1997 J. Math. Phys. {\bf 38}  6675

\bibitem{RCP98} Rigo A, Casas M and Plastino A 1998 Phys. Rev A {\bf 57}
 2319.

\bibitem{HH96}  Horodecki R and Horodecki M 1996 Phys. Rev. A {\bf 54}
 1838.

\bibitem{P96} Peres A 1996 Phys. Rev. Lett. {\bf 77}  1413

\bibitem{HHH96} Horodecki M, Horodecki P and Horodecki R
1996 Phys. Lett. A {\bf 223} 1.

\end{thebibliography}
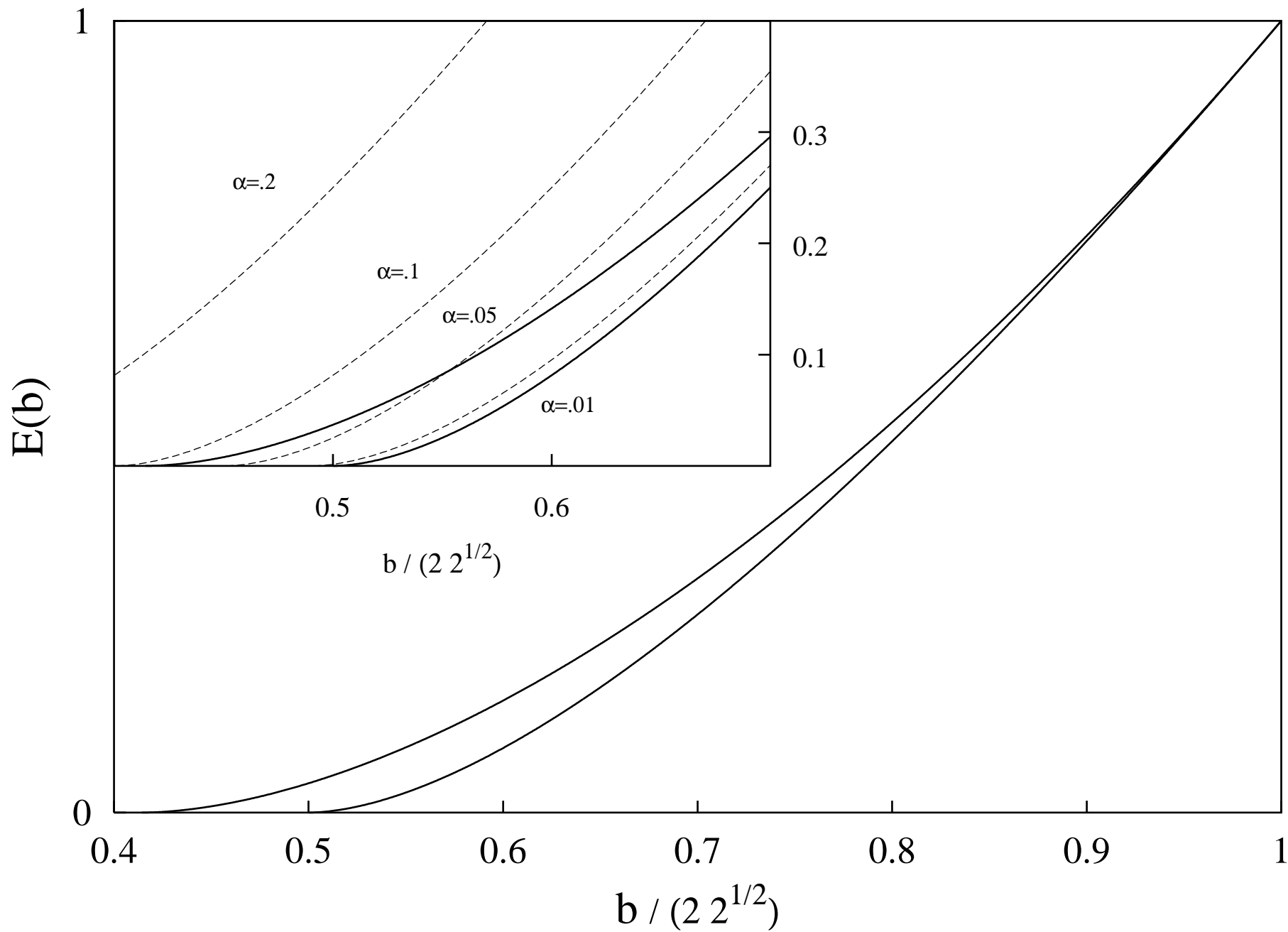
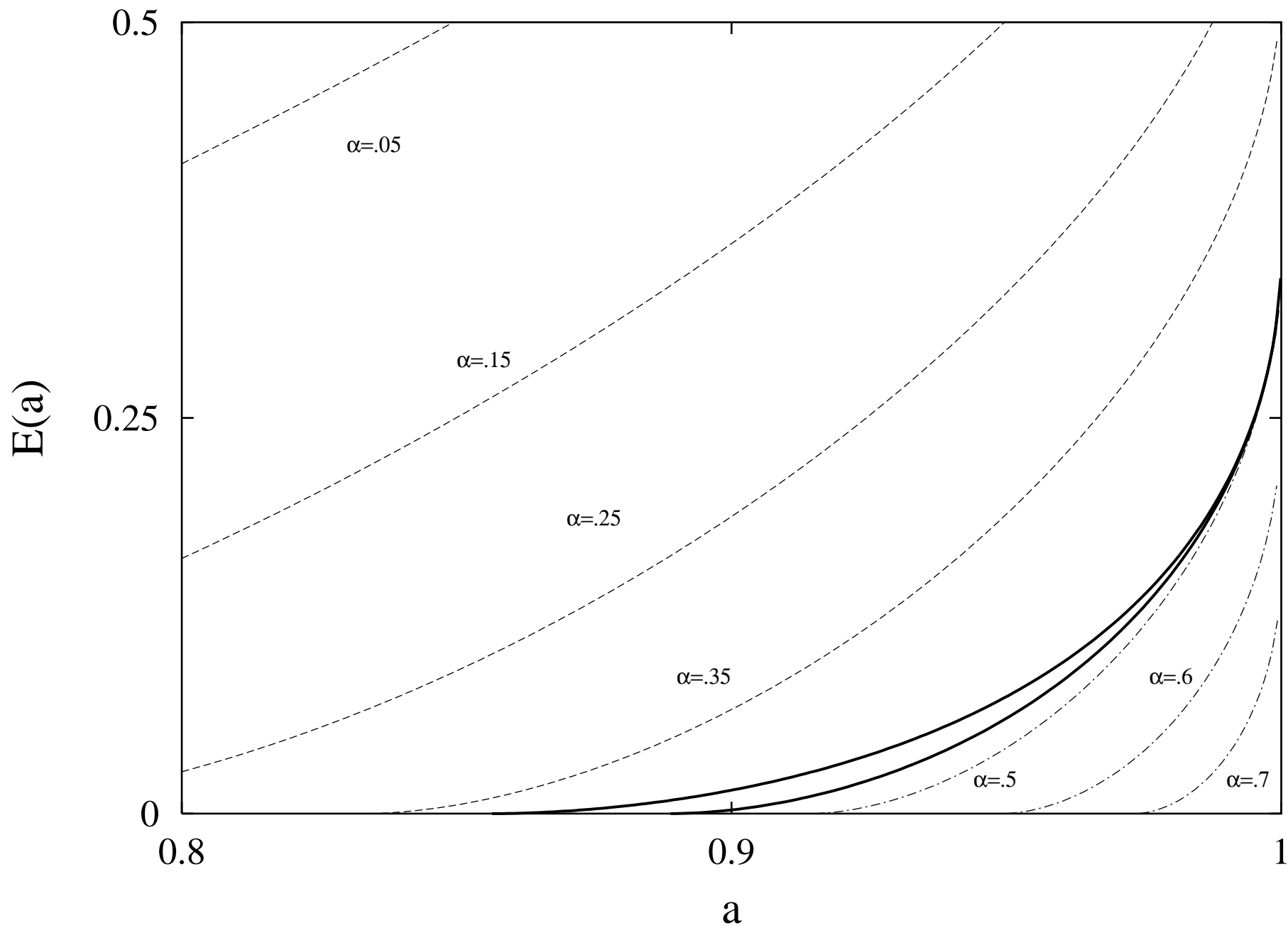
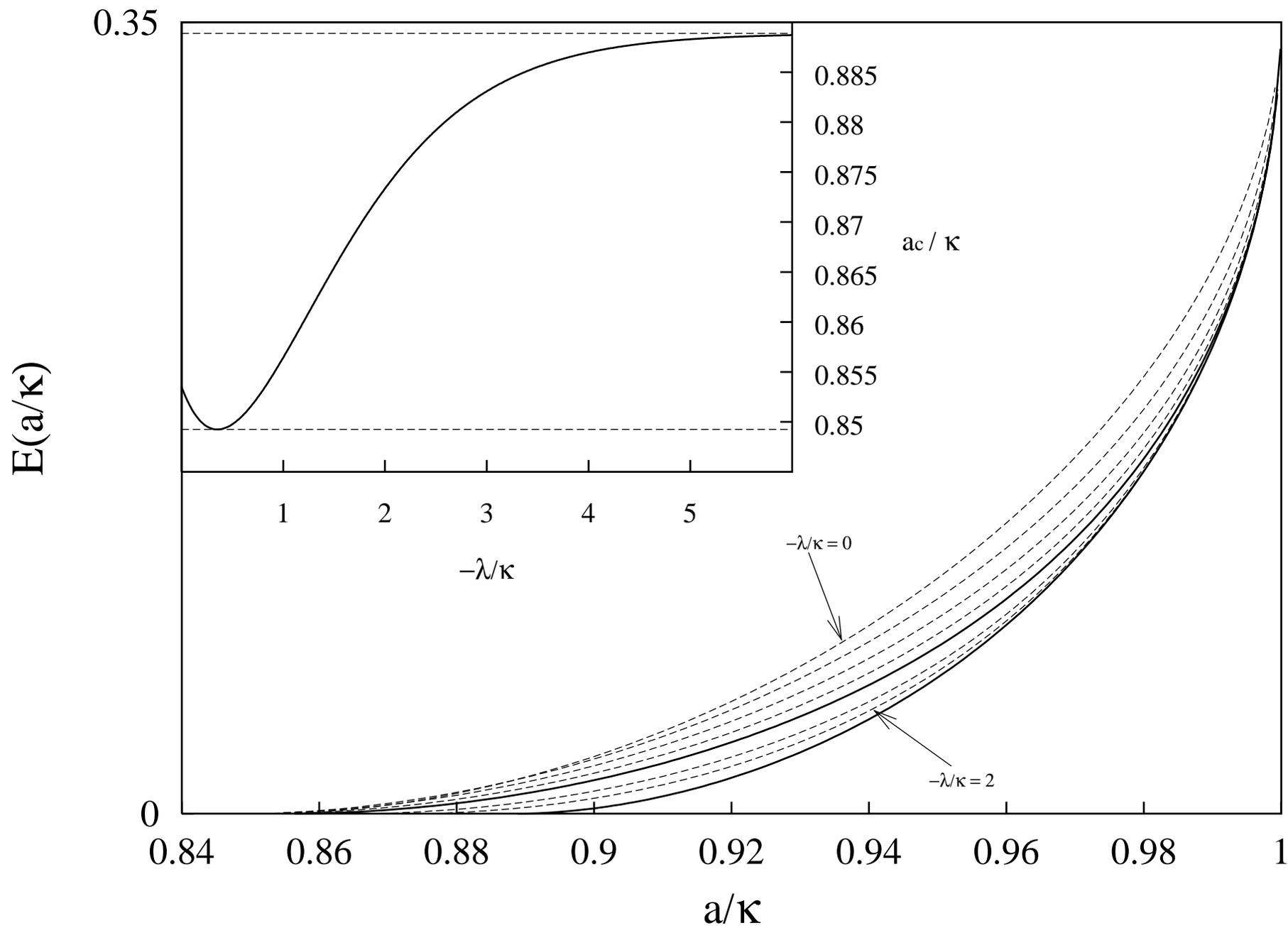
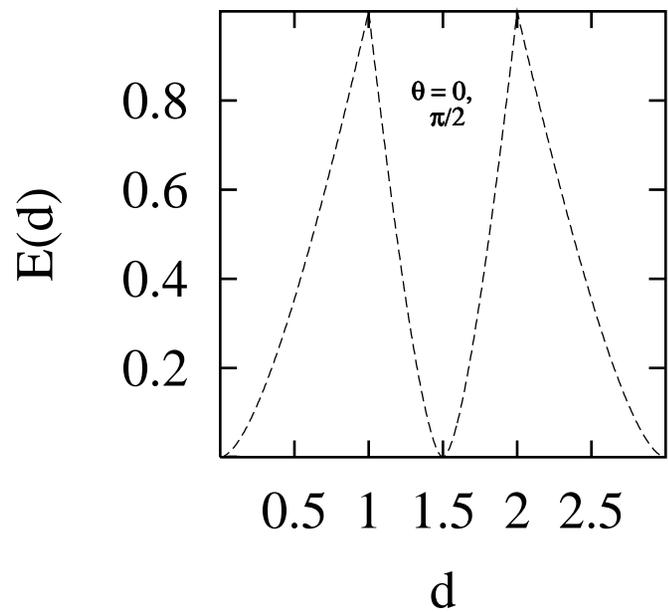
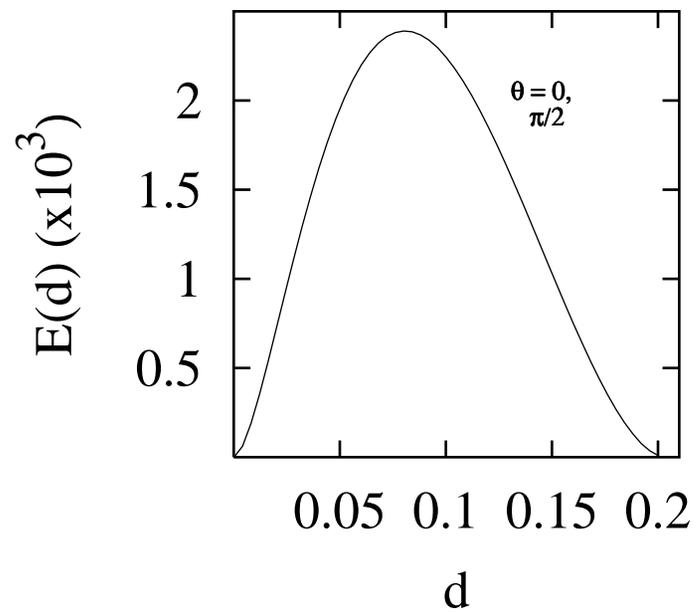
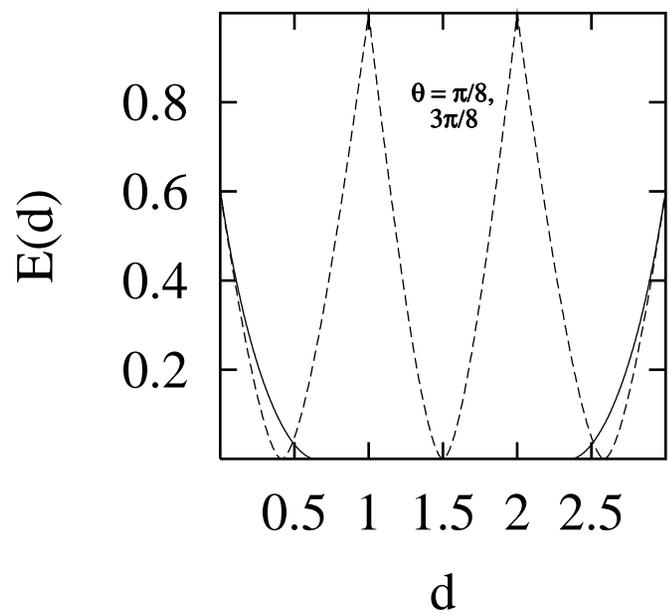
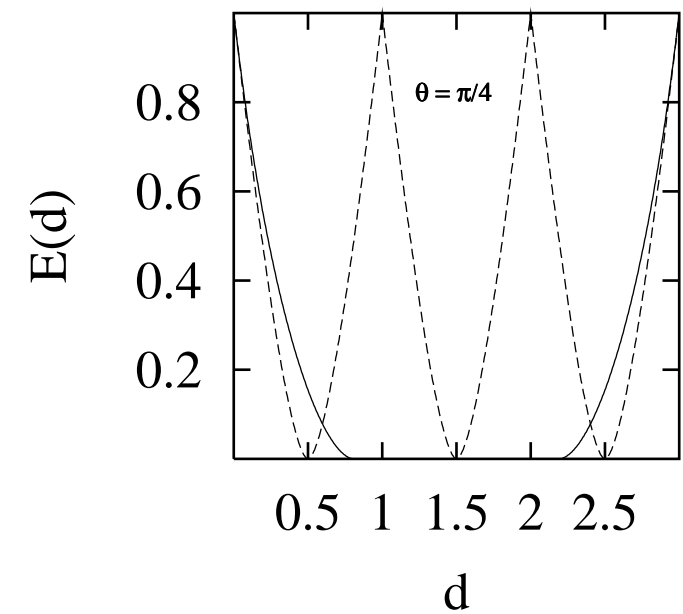
\end{document}